\documentstyle[amssymb,12pt,preprint,aps]{revtex}
\begin{document}
\title{Initial Conditions for Supersymmetric Inflation}
\author{{\bf G. Lazarides}}
\address{Physics Division\\
School of Technology\\
University of Thessaloniki\\
Thessaloniki 540 06, Greece}
\author{{\bf N.D.Vlachos}}
\address{Dept. of Theoretical Physics\\
University of Thessaloniki\\
Thessaloniki 540 06, Greece }
\maketitle

\begin{abstract}
We perform a numerical investigation of the fields evolution in the
supersymmetric inflationary model based on radiative corrections.
Supergravity corrections are also included. We find that, out of all the
examined initial data, only about 10\% give an adequate amount of inflation 
and can be considered as ''natural''.
Moreover, these successful initial conditions appear scattered and more 
or less isolated.
\end{abstract}

\newpage

A couple of years ago Linde$\cite{Linde}$ has proposed, in the context of
non-supersymmetric (SUSY) grand unified theories (GUTs), a clever
inflationary scenario based on a coupled system of two real scalar fields
one of which is a gauge non-singlet. This field remains displaced from its
vacuum value during inflation and provides the vacuum energy which drives
inflation while the other, which is a gauge singlet, is the slowly varying
field during inflation. The main advantage of this hybrid inflationary
scenario is that it can reproduce, in contrast to previous realizations of
inflation, the observed temperature fluctuations in the cosmic background
radiation (CBR) with natural values of the coupling constants. However,
inflation terminates abruptly and is followed by a ''waterfall'' regime
during which topological defects are easily produced if they are predicted
by the symmetry breaking associated with inflation.

Recently \cite{LazPan,Shafi}, two variants of the above scenario,
both sharing its naturalness, were proposed in the context of SUSY GUTs. One
of them \cite{LazPan} relies on the lowest order non-renormalizable
contribution to the superpotential. 
This scenario has been called smooth hybrid inflation
since the termination of the inflationary period is not as abrupt as in
Linde's scheme. Rather, the entrance of the system into the oscillatory phase
is quite smooth. Moreover, the system, already from the beginning of
inflation, follows a particular valley of minima that leads to a particular
point of the vacuum manifold and, thus, there is no production of
topological defects in this case. This is certainly an advantage of this
scenario since it makes it applicable even to GUT models where the symmetry
breaking associated with inflation does imply the existence of (possibly)
cosmologically disastrous topological defects like domain walls or magnetic
monopoles. The main advantage though of smooth hybrid inflation is that it
can be associated to the SUSY GUT\ scale $M_{SG}\simeq 2\times 10^{16}$ GeV
consistent with the unification of the minimal supersymmetric standard model
(MSSM) gauge couplings. A possible shortcoming of this scenario is that the
relevant part of inflation takes place at relatively high values 
[$(1-3)\times 10^{17}$ GeV] of the inflaton field and, thus, supergravity
and superstring corrections may not be so easily tamable.

The second SUSY variant \cite{Shafi} of the hybrid inflationary
scenario utilizes the one loop radiative corrections to the inflationary
potential. These corrections are non-trivial since SUSY is broken at the
inflationary trajectory in field space. The termination of inflation is as
in the non-SUSY Linde's scenario and topological defects can be copiously
produced. It is then important to make sure that no cosmologically dangerous
topological defects come into existence as a result of the symmetry breaking
associated to inflation. Moreover, as it turns out, the scale at which this
symmetry breaking takes place is about five times smaller than $M_{SG}$ .
This can be considered as a disadvantage of the scheme since it forces us to
abandon the successful unification of the MSSM gauge coupling constants.
In this case, inflation may be chosen to take place during an intermediate
symmetry breaking with no magnetic monopoles or domain walls associated with
it. The important advantage of this scenario is that the relevant part of
inflation takes place at values of the inflaton field of order
$10^{16}$ GeV which are low enough for supergravity and superstring
corrections to be under relatively good control \cite{LazSSh}.

For an inflationary scenario to be considered fully successful, one has to
show that it is obtainable for a wide class of ''natural'' initial values of
the fields and their time derivatives. In ref. \cite{LaPanV} 
this issue was investigated for smooth hybrid inflation.
The result was striking and unexpected. 
It was found that, for almost all the examined initial conditions
(except for a narrow transition region), we get inflation with an adequate
number of e-foldings. This is certainly an important advantage of this
inflationary scenario. The purpose of this work is to discuss the question
of initial conditions for the second SUSY hybrid inflationary
scenario which is based on radiative corrections. In particular, we will try
to see if there exists a wide class of "natural" initial conditions, i.e.,
a class of comparable initial values of the fields with vanishing 
time derivatives for which the system
falls at the bottom of the valley of minima so that its subsequent evolution
along this valley produces an adequate amount of inflation. To this end, we
solve numerically the evolution equations of the system for various sets of
initial conditions. Our analysis also includes cases with all initial field
values being much smaller than the Planck scale, where the results are
expected to be less affected by replacing global by local SUSY. At
the end, non-vanishing initial time derivatives of the fields
as well as supergravity corrections were also taken into account.

The SUSY inflationary scenario based on radiative corrections can
be realized in the context of a SUSY GUT model with a (semisimple) gauge
group $G$ of rank $\geq 5$ which breaks spontaneously directly to the
standard model (SM) group $G_{S}$ at a scale $M$. The symmetry breaking of $G
$ to $G_{S}$ is obtained through a superpotential which includes the terms 
\cite{inflsup} 
\begin{equation}
W=\kappa S(-M^{2}+\,\bar{\phi}\phi ).  \label{susypot}
\end{equation}
Here, $\bar{\phi}$, $\phi$ is a conjugate pair of left-handed SM singlet
superfields which belong to non-trivial representations of the gauge group $G
$ and reduce its rank through their vacuum expectation values (vevs) and $S$
is a gauge singlet left-handed superfield. The superpotential in eq.(\ref
{susypot}) is in fact the only renormalizable superpotential consistent
with a continuous $U(1)$ R-symmetry under which $W\rightarrow e^{i\theta }W$
, $S\rightarrow e^{i\theta }S$, $\bar{\phi} \phi\rightarrow \bar{\phi} \phi$
. The potential obtained from $W$ in eq.(\ref{susypot}), in the
SUSY limit, is 
\begin{equation}
V=\kappa ^{2}\mid M^{2}-\bar{\phi}\phi \mid ^{2}+\kappa ^{2}\mid S\mid
^{2}(\mid \bar{\phi} \mid ^{2}+\mid \phi\mid ^{2})+{\textstyle{D-terms}{
\quad \cdot }}  \label{susypot1}
\end{equation}
Vanishing of the D-terms is achieved along the D-flat directions where 
$|\bar{\phi}|=|\phi |$ . The SUSY vacuum 
\begin{equation}
<S>=0,\>\><\bar{\phi} ><\phi>=M^{2},\>\>|<\bar{\phi}>|=|<\phi >|
\end{equation}
lies on the particular D-flat direction $\bar{\phi}^{*}=\phi $. Restricting
ourselves to this direction and performing appropriate gauge and
R-transformations we can bring the complex $S$,$\bar{\phi}$,$\phi$ fields
on the real axis, i.e., $S\equiv \frac{\sigma }{\sqrt{2}}$, $\>\bar{\phi}
=\phi \equiv \frac{1}{2}\chi $, where $\sigma $ and $\chi $ are real scalar
fields. The potential in eq.(\ref{susypot1}) then takes the form 
\begin{equation}
V(\chi ,\sigma )=\kappa ^{2}(M^{2}-\frac{\chi ^{2}}{4})^{2}+\frac{\kappa
^{2}\chi ^{2}\sigma ^{2}}{4} \quad \cdot \label{scpot}
\end{equation}
The vacua lie at $<\chi >=\pm 2M$, $<\sigma >=0$. We observe that the
potential possesses an exact flat direction at $\chi =0$ with $V(\chi
=0,\sigma )=\kappa ^{2}M^{4}$. The mass squared of the field $\chi $ along
this flat direction is given by $m_{\chi }^{2}=-\kappa ^{2}M^{2}+\frac{1}{2}
\kappa ^{2}\sigma ^{2}$ and remains non-negative for $\sigma \geq \sigma
_{c}=\sqrt{2}\,M$. This implies that, at $\chi =0$ and $\sigma \geq \sigma
_{c}$, we obtain a flat-bottomed valley of minima which, as it stands, is
not suitable for inflation. However, due to SUSY breaking along the bottom
of this valley resulting from the non-zero value of $V(\chi =0,\sigma )$,
there exist non-trivial radiative corrections to the effective potential so
that this valley acquires the negative slope which is necessary for
inflation. The one loop corrected effective potential (along the direction 
$\sigma \geq \sigma _{c}\,,\chi =0$ ) is given by \cite{Shafi} 
\begin{equation}
\begin{array}{c}
V_{eff}(\sigma)=\kappa ^{2}M^{4}+\displaystyle{\frac{\kappa ^{4}}{32\pi ^{2}}}
[2M^{4}\ln\displaystyle{\frac{\kappa ^{2}\sigma ^{2}}
{2\Lambda ^{2}}}+\left( \frac{1}{2}\,\sigma
^{2}-M^{2}\right) ^{2}\,\ln \left( 1-\frac{2M^{2}}{\,\sigma ^{2}}\right) +
\\ 
\displaystyle{\left( \frac{1}{2}\sigma ^{2}+M^{2}\right) ^{2}\,
\ln \left( 1+\frac{2M^{2}}{\sigma ^{2}}\right) ]\, \cdot}
\end{array}
\label{effpot1}
\end{equation}

\noindent If $\sigma \gg \sigma _{c}$, $V_{eff}(\sigma )$ in eq.(\ref{effpot1}) 
reduces to the simpler form 
\begin{equation}
V_{eff}(\sigma \gg \sigma _{c})\approx \kappa ^{2}M^{4}\left[ 1+\frac{\kappa
^{2}}{16\pi ^{2}}\left( \ln \frac{\kappa ^{2}\,\sigma ^{2}}{2\Lambda ^{2}}+
\frac{3}{2}\right) \right] \quad \cdot   \label{reducv}
\end{equation}

A region of the universe, where $\chi $ and $\sigma $ happen to be almost
uniform with negligible kinetic energies and with values close to the bottom
of the valley of minima, follows this valley in its subsequent evolution and
undergoes inflation. The temperature fluctuations of CBR produced during
this inflation can be estimated to be \cite{Shafi} 
\begin{equation}
\frac{\delta T}{T}\simeq \sqrt{\frac{N_{\text{{\sc q}}}}{45}}\left( \frac{M}{
M_{\text{c}}}\right) ^{2},  \label{dtovt}
\end{equation}
where $M_{\text{c}}=M_{\text{{\sc p}}}/\sqrt{8\pi }$, $M_{\text{{\sc p}
}}=1.22\times 10^{19}$ GeV is the Planck mass, and $N_{\text{{\sc q}}}\simeq
60$ is the number of e-foldings of our present horizon scale during
inflation. The Cosmic Background Explorer (COBE) result, $\delta T/T\simeq
6.6\times 10^{-6}$, can then be reproduced with $M=5.82\times 10^{15}$ GeV.
Using eq.(\ref{reducv}) one finds 
\begin{equation}
\kappa =\pi \left( \frac{8}{N_{\text{{\sc q}}}}\right) ^{1/2}\frac{\sigma _{
{\sc Q}}}{\sigma _{\text{c}}}\frac{M}{M_{\text{c}}}\quad ,
\label{kappa}
\end{equation}
where $\sigma _{\text{{\sc q}}}$ is the value of $\sigma $ when the scale
which evolved to our present horizon size crossed outside the de Sitter
horizon during inflation. Inflation continues until $\sigma =\sigma _{\text{
c}}$ where it terminates abruptly and is followed by a ''waterfall'',
i.e., a sudden entrance into an oscillatory phase about a global minimum. This
is checked by noting the magnitude of the quantities $\epsilon =\left(
V^{\prime }/V\right) ^{2}M_{\text{c}}^{2}/2,\eta =M_{\text{c}
}^{2}\,V^{\prime \prime }/V$, where the prime refers to derivatives with
respect to $\sigma $ . Since the system can fall into either of the two
available global minima with equal probability, topological defects can be
easily produced if they are predicted by the relevant symmetry breaking 
of the particular particle physics model one is employing.

We will now try to specify the initial conditions for the $\sigma $ and 
$\chi $ fields which lead to the above described inflationary scenario. In
other words, we will try to identify initial conditions for which the system
falls at the bottom of the valley of minima of the effective potential so
that its subsequent evolution along this valley produces an adequate amount
of inflation. We assume that, after ''compactification'' at some initial
cosmic time, a region emerges in the universe where the scalar fields $
\sigma $ and $\chi $ and their time derivatives happen to be 
almost uniform and the energy density $\lesssim M_{c}^{4}$. 
(The initial values of $\sigma $ and $\chi$ together 
with their time derivatives can always
be transformed by appropriate gauge and R-transformations to become
positive). The radiative corrections in eqs.(\ref{effpot1}) and (\ref{reducv}),
though crucial for inflation, play no essential role in the issue of initial
conditions and, thus, they will be ignored. In order to keep the formalism
as clear as possible, we shall from now on use dimensionless variables for
all relevant quantities as follows: 
\begin{equation}
\hat{\chi}=\frac{\chi }{M_{\text{c}}}\ ,\hat{\sigma}=\frac{\sigma }{M_{
\text{c}}},\ \hat{t}=\kappa M_{\text{c}}t,\ \hat{M}=\frac{M}{M_{
\text{c}}},\ \hat{H}=\frac{H}{\kappa M_{\text{c}}},\ \hat{\varrho}=
\frac{\varrho }{\kappa ^{2}M_{\text{c}}^{4}},\ \hat{V}(\hat{\chi},\hat{
\sigma})=\frac{V(\chi ,\sigma )}{\kappa ^{2}M_{\text{c}}^{4}} ,
\label{resclq}
\end{equation}
where $H$ is the Hubble parameter and $\varrho $ the energy density.
The evolution of the system in this region is governed by the
following equations of motion 
\begin{equation}
\ddot{\hat{\chi}}+3\hat{H}\dot{\hat{\chi}}-\,\hat{\chi}\,(\hat{M}^{2}-\frac{
\hat{\chi}^{2}}{4})+\frac{1}{2}\hat{\sigma}^{2}\hat{\chi}=0\ ,  \label{xevol}
\end{equation}
\begin{equation}
\ddot{\hat{\sigma}}+3\hat{H}\dot{\hat{\sigma}}+\frac{1}{2}\hat{\chi}^{2}\hat{
\sigma}=0\ ,  \label{sevol}
\end{equation}
where overdots denote derivatives with respect to the dimensionless time 
$\hat{t}$, the rescaled Hubble parameter $\hat{H}$ is given by 
\begin{equation}
\hat{H}=\frac{\hat{\varrho}^{1/2}}{\sqrt{3}}=\frac{1}{\sqrt{3}}\left( \frac{1
}{2}\dot{\hat{\chi}}^{2}+\frac{1}{2}\dot{\hat{\sigma}}^{2}+\hat{V}(\hat{\chi}
,\hat{\sigma})\right) ^{1/2},\quad \hat{V}(\hat{\chi},\hat{\sigma})=(\hat{M}
^{2}-\frac{\hat{\chi}^{2}}{4})^{2}+\frac{\hat{\chi}^{2}\hat{\sigma}^{2}}{4}
\label{reschub}
\end{equation}
and $\hat{M}=2.39\times 10^{-3}$.

We shall first examine the case where initially $\hat{\sigma}\gg \hat{\chi},1$.
Under these conditions, the last term in eq.(\ref{xevol}) dominates and
this equation reduces to 
\begin{equation}
\ddot{\hat{\chi}}+3\hat{H}\dot{\hat{\chi}}+\frac{1}{2}\hat{\sigma}^{2}
\hat{\chi}\simeq0\quad .  \label{xevol1}
\end{equation}
Furthermore, the potential energy density $\hat{V}(\hat{\chi},\hat{\sigma})$
in eq.(\ref{reschub}) is dominated by its last term provided that $\hat{\chi}
\hat{\sigma}\gg 2\hat{M}^{2}$. Let us for the moment assume that, during the
early stages of evolution, $\hat{\sigma}$ remains almost constant.
Then, for $|\dot{\hat{\chi}}|\ll \hat{\chi}\hat{\sigma}/\sqrt{2}$,
$\hat{H}$ is given approximately by 
\begin{equation}
\hat{H}\simeq \frac{1}{2\sqrt{3}}\hat{\chi}\hat{\sigma}\ ,  \label{h1}
\end{equation}
so that $\hat{H}$ and $\hat{\chi}$ satisfy the following differential
equations 
\begin{equation}
\ddot{\hat{\chi}}+\frac{\sqrt{3}}{2}\hat{\sigma}\hat{\chi}
\dot{\hat{\chi}}+\frac{1}{2}
\hat{\sigma}^{2}\hat{\chi}\simeq 0\ ,  \label{xevapr}
\end{equation}
\begin{equation}
\ddot{\hat{H}}+3\hat{H}\dot{\hat{H}}+\frac{1}{2}\hat{\sigma}^{2}\hat{H}
\simeq 0\ \cdot 
\end{equation}
Integrating once, with $\dot{\hat{\chi}}(\hat{t}=0)=0$ for 
simplicity, we get 
\begin{equation}
\hat{\chi}(\hat{t})^{2} \simeq \hat{\chi}(0)^{2}+\frac{4}{3}\ln 
\left[ 1+\frac{\sqrt{3}}{\hat{\sigma}}\dot{\hat{\chi}}
(\hat{t})\right] -\frac{4}{\sqrt{3}\hat{
\sigma}}\dot{\hat{\chi}}(\hat{t})\ ,
\end{equation}
\begin{equation}
\hat{H}(\hat{t})^{2} \simeq \hat{H}(0)^{2}+\frac{\hat{\sigma}^{2}}{9}
\ln \left[ 1+\frac{6}{\hat{\sigma}^{2}}\dot{\hat{H}}(\hat{t})\right]
 -\frac{2}{3}\dot{\hat{H}}(\hat{t})\ \cdot   \label{hevol}
\end{equation}
Using these equations one can show that $\hat{\chi}(\hat{t})$, 
$\dot{\hat{\chi}}(\hat{t})$,
 $\hat{H}(\hat{t})$ and $\dot{\hat{H}}(\hat{t})$ are initially
(for $\hat{\chi}(\hat{t})>0$) monotonically decreasing functions of
$\hat{t}$. At some time $\hat{t}_{0}$, $\hat{\chi}(\hat{t})$
becomes small enough and starts performing damped oscillations over
the maximum at $\hat{\chi}=0$.
For $\hat{t}\geq \hat{t}_0$, the continuity equation 
$
\dot{\hat{\varrho}}=-3\hat{H}(\hat{\varrho}+\hat{p})\ ,
$
with $\hat{p}=p/\kappa ^{2}M_{c}^{4}$, $p$ being the pressure, 
averaged over one oscillation of $\hat{\chi}$ becomes 
$
\dot{\hat{\varrho}}=-3\hat{H}\gamma \hat{\varrho}\ ,
$
where $\gamma =1$ for a $\hat{\chi}^{2}$ potential$\cite{Turner}$. This
together with the fact that $\hat{\varrho}$ is proportional to $
\hat{H}^{2}$ gives 
\begin{equation}
\hat{H}\simeq \frac{2}{3\hat{t}^{\prime }},\quad \hat{t}^{\prime }\equiv 
\hat{t}-\hat{t}_{0}+\frac{2}{3}\hat{H}(\hat{t}_{0})^{-1}, 
\hat{t}\geq \hat{t}_0 \quad \cdot 
\label{hubest}
\end{equation}
The quantities $\hat{\chi}(\hat{t}_{0})\,,\hat{H}(\hat{t}_{0})$ as well as
their time derivatives can be estimated as follows. When the $\hat{\chi}$
field starts oscillating, $\hat{H}(\hat{t})$ satisfies the equation 
\begin{equation}
\hat{H}(\hat{t}_{0})^{2}\simeq -\frac{2}{3}\dot{\hat{H}}(\hat{t}_{0})\quad , 
\label{hhdot}
\end{equation}
as one deduces from eq.(\ref{hubest}). Eqs.(\ref{h1}) and 
(\ref{hevol}) then give 
\begin{equation}
\dot{\hat{\chi}}(\hat{t}_{0}) \simeq \frac{2\sqrt{3}}{
\hat{\sigma}}\dot{\hat{H}}(\hat{t}
_{0})\simeq -\frac{\hat{\sigma}}{\sqrt{3}}\left[ 
1-\exp [-\frac{3}{4}\hat{\chi}
(0)^{2}]\right] \cdot  \label{incond1}
\end{equation}
Substituting into eq.(\ref{hhdot}) we, furthermore, get 
\begin{equation}
\hat{\chi}(\hat{t}_{0}) \simeq \frac{2\sqrt{3}}{\hat{\sigma}}
\hat{H}(\hat{t}_{0})
\simeq \frac{2}{\sqrt{3}}\sqrt{1-
\exp [-\frac{3}{4}\hat{\chi}(0)^{2}]}\quad \cdot 
\label{incond2}
\end{equation}
Note that eqs.(\ref{incond1}) and (\ref{incond2}) imply that 
$|\dot{\hat{\chi}}|<
\hat{\chi}\,\hat{\sigma}/\sqrt{2}$, for $\hat{t}\leq \hat{t}_{0}$,
as was required for the validity of eq.(\ref{h1}). 
Also, for $\hat{t}\leq \hat{t}_{0}$,
$\hat{\sigma}\gg \hat{\chi}\gtrsim 1$ and in eq.(\ref{sevol}) the
Hubble parameter dominates over the ''frequency'' $\hat{\chi}/\sqrt{2}$ .
This equation then reduces to $\dot{\hat{\sigma}}$ $\thickapprox -
\hat{\chi}/\sqrt{3}$
and the variation of $\hat{\sigma}$ till $\hat{t}_{0}$ is of order $\hat{
\chi}(0)\hat{H}(\hat{t}_{0})^{-1}\thicksim \hat{\chi}(0)/\hat{\sigma}\ll 1$.
This justifies our assumption that $\hat{\sigma} $ remains constant for
$\hat{t}\leq \hat{t}_{0}.$

Upon substituting eq.(\ref{hubest}) into eq.(\ref{xevol1}) we get 
\begin{equation}
\ddot{\hat{\chi}}+\frac{2}{\hat{t}^{\prime }}\dot{\hat{\chi}}+
\frac{1}{2}\hat{\sigma}^{2}
\hat{\chi}\simeq 0\quad \cdot 
\end{equation}
For $\hat{\sigma}$ constant, this is a differential equation of the Emden
type whose solution satisfying the (approximate) boundary conditions at $
\hat{t}_{0}$ is given by 
\begin{equation}
\hat{\chi}(\hat{t}) \simeq \frac{4}{\sqrt{3\,}\,
\hat{\sigma}\hat{\,t}^{\prime }}
\cos \left[ \frac{1}{\sqrt{2}}\hat{\sigma}\,(\hat{t}-\hat{t}_{0})\right]
,\quad \hat{t}\geq \hat{t}_{0}\cdot   \label{xfin}
\end{equation}
Eq.(\ref{sevol}), for $\hat{t}\gg \hat{t}_{0}$, averaged over one
oscillation of $\hat{\chi}$ then gives 
\begin{equation}
\ddot{\hat{\sigma}}+\frac{2}{\hat{t}}\dot{\hat{\sigma}}+
\frac{4}{3\hat{\sigma}\,\hat{t}
^{2}}\simeq 0\ \cdot 
\end{equation}
This equation implies 
\begin{equation}
\hat{t}-\hat{t}_{0} \simeq \displaystyle\int_{x(
\hat{t}_{0})}^{x(\hat{t})}\frac{dx}{
\sqrt{-\frac{4}{3}\ln \displaystyle \frac{x^{2}}{x(\hat{t}_{0})^{2}}+\dot{x}(
\hat{t}_{0})^{2}}}\quad ,   \label{sol}
\end{equation}
where $x(\hat{t})\equiv \hat{t}\hat{\sigma}(\hat{t})$,
$\dot{x}(\hat{t})>0$.We see
that $x(\hat{t})$ is a periodic function of time corresponding to a bounded
motion of $x$ in a logarithmic potential. The turning point $x_{m}(>0)$ is
defined through the relation 
\begin{equation}
\frac{4}{3}\ln x(\hat{t}_{0})^{2}+\dot{x}(\hat{t}_{0})^{2}=\frac{4}{3}\ln
x_{m}^{2}  \label{turnp}
\end{equation}
and eq.(\ref{sol}) becomes ($x(\hat{t})>0$) 
\begin{equation}
\hat{t}-\hat{t}_{0}\simeq \frac{\sqrt{3\pi }}{2\sqrt{2}}x_{m}
\left[ \Phi (\sqrt{
\ln \frac{x_{m}}{x(\hat{t}_{0})}})-\Phi (\sqrt{\ln 
\frac{x_{m}}{x(\hat{t})}})\right] \ ,  \label{tsol}
\end{equation}
where $\Phi (x)$ is the probability integral. The turning point $x_{m}$ can
be estimated from eq.(\ref{turnp}) and the fact that
$\dot{x}(\hat{t}_{0})=\dot{\hat{
\sigma}}(\hat{t}_{0})\hat{t}_{0}+\hat{\sigma}(\hat{t}_{0})\thickapprox \hat{
\sigma}(0)$. We get $x_{m}\thickapprox x(\hat{t}_{0})\exp (3\hat{\sigma}
(0)^{2}/8)\gg x(\hat{t}_{0})$ for $\hat{\sigma}(0)\gg 1$. This implies that
the system,at $\hat{t}=\hat{t}_{0}$, starts from a position far
away from the turning point which could subsequently be reached after the
lapse of time $\delta \hat{t}\thickapprox x_{m}\sqrt{3\pi }/2\sqrt{2}
\thickapprox (\sqrt{3\pi }/2\sqrt{2})\exp (3\hat{\sigma}(0)^{2}/8)x(\hat{t}
_{0})$ as one deduces from eq.(\ref{tsol}).Using $x(\hat{t}_{0})=\hat{\sigma}
(\hat{t}_{0})\hat{t}_{0}\thickapprox (2/3)\hat{\sigma}(0)\hat{H}(\hat{t}
_{0})^{-1}\thickapprox 2[1-\exp [-3\hat{\chi}(0)^{2}/4]]^{-1/2}$ from eq.(\ref
{incond2}), $\delta \hat{t}$ can be related to the initial values of the
fields $\hat{\chi}(0)$, $\hat{\sigma}(0)$. The time needed for the amplitude
of $\hat{\chi}$ to drop to about $2\hat{M}^{2}/\hat{\sigma}$, where the $
\hat{M}^{4}$ term starts dominating the potential energy $\hat{V}(\hat{\chi},
\hat{\sigma})$ in eq.(\ref{reschub}), can be estimated from eq.(\ref{xfin})
to be $\hat{t}_{\text{{\sc d}}}\thickapprox 2/\sqrt{3}\hat{M}^{2}\thickapprox
2\cdot 10^{5}\ll \delta \hat{t}$ for $\hat{\sigma}(0)\gg 1$ and $\hat{\chi}
(0)$ not extremely small. We then deduce that $\hat{\sigma}$ is an extremely
slowly decreasing function of time, for $\hat{t}_{0}$ $\leq \hat{t}
\leq \hat{t}_{\text{{\sc d}}}$ ,
and, thus, our assumption that $\hat{\sigma}$ remains
constant is justified in this time interval too.

For cosmic times $\geq \hat{t}_{\text{{\sc d}}}$, the $\hat{M}^{4}$ 
term dominates
the potential energy in eq.(\ref{reschub}) and the Hubble parameter becomes
approximately constant and equal to $\hat{H}=\hat{M}^{2}/\sqrt{3}$ and
remains so thereafter. Eq.(\ref{xevol1}) still holds and, assuming that $
\hat{\sigma}$ remains essentially unchanged, $\hat{\chi}$ performs damped
oscillations of frequency $\hat{\sigma}/\sqrt{2}\gg \hat{H}$ . Eq.(\ref
{sevol}) averaged over one oscillation of $\hat{\chi}$ then becomes 
\begin{equation}
\ddot{\hat{\sigma}}+3\hat{H}\dot{\hat{\sigma}}+\frac{1}{4}\hat{\chi}_{m}^{2}
\hat{\sigma}=0\,\ ,  \label{sevol2}
\end{equation}
where $\hat{\chi}_{m}$ is the amplitude of $\hat{\chi}$. The decreasing
''frequency'' of $\hat{\sigma}$ which is $\hat{\chi}_{m}/2\leq \hat{M}^{2}/
\hat{\sigma}$ soon becomes much smaller than $\hat{H}$ and eq.(\ref{sevol2})
reduces to 
\begin{equation}
3\hat{H}\dot{\hat{\sigma}}+\frac{1}{4}\hat{\chi}_{m}^{2}
\hat{\sigma}\thickapprox 0 
\quad \cdot
\end{equation}
The variation of $\hat{\sigma}$ within one expansion time is then estimated
to be $|\Delta \hat{\sigma}/\hat{\sigma}|\thicksim \hat{\chi}_{m}^{2}/12\leq 
\hat{M}^{4}/3\hat{\sigma}^{2}\ll 1$. This justifies the assumption that $
\hat{\sigma}$ does not change much. The overall conclusion is that, at times
much larger than 
$\hat{t}_{\text{{\sc d}}}\thickapprox 2/\sqrt{3}\hat{M}^{2}$, the 
$\hat{\chi}$ field falls into the valley of minima of the potential in eq.(
\ref{reschub} ) and relaxes at the bottom of this valley 
whereas the $\hat{\sigma}$ field still remains more or less unchanged.
After that, the radiative
corrections in eq.(\ref{effpot1}) come into play and the system follows the
valley of minima towards the supersymmetric vacuum and, therefore, the
hybrid inflationary scenario is realized for initial values of the fields
satisfying the inequality $\hat{\sigma} \gg 1$, $\hat{\chi} $. These 
initial conditions, however, cannot be considered as totally 
satisfying because there is a
considerable discrepancy between the initial values of the fields. Also, the
inclusion of supergravity is expected to invalidate the above discussion of
initial conditions which involve values of the field $\hat{\sigma} \geq 1$. 

For SUSY inflation based on radiative corrections to be considered as a
fully successful inflationary scenario, one must show that it is obtained
for a wide class of initial conditions which are more "natural" than the
ones just discussed. This can be checked only numerically. To this end, we
have conducted numerical integration of eqs.(\ref{xevol}) and (\ref{sevol})
for the following sets of initial conditions:

\begin{enumerate}
\item  $4\geq \hat{\chi}\geq 0.5$, $6\geq \hat{\sigma}\geq 0.5$ with
vanishing initial velocities.

\item  $1\geq \hat{\chi}\geq 0.1$, $1\geq \hat{\sigma}\geq 0.1$ with
vanishing initial velocities.

\item  $1\geq \hat{\chi}\geq 0.1$, $1\geq \hat{\sigma}\geq 0.1$ at the same
points as in (2) introducing equal positive initial velocities so that the
initial kinetic energy equals the corresponding potential energy.
\end{enumerate}
The last two cases, although corresponding to pretty low values of the
initial energy density, were included since our results there are expected to
remain essentially unaffected if we replace global by local SUSY
with canonical Kaehler potential. On the contrary, the results of case (1)
should be strongly affected by supergravity corrections. To see this we
reexamined this case by including supergravity corrections with 
minimal Kaehler potential. This is done by replacing the 
potential energy density in eq.(\ref{reschub}) and its derivatives with
respect to $\hat{\chi}$ and $\hat{\sigma}$ in eqs.(\ref{xevol}) and 
(\ref{sevol}) by 
\[
\hat{V}(\hat{\chi},\hat{\sigma})_{\text{{\sc sg}}}=\exp \left[ \frac{1}{2}\hat{
\chi}^{2}+\frac{1}{2}\hat{\sigma}^{2}\right] \times 
\]
\begin{equation}
\left\{ \left( \frac{1}{4}\hat{\chi}^{2}-\hat{M}^{2}\right) ^{2}\left[ 1-
\frac{1}{2}\hat{\sigma}^{2}+\frac{1}{4}\hat{\sigma}^{2}(\hat{\chi}^{2}+\hat{
\sigma}^{2})\right] +\frac{1}{2}\hat{\sigma}^{2}\hat{\chi}^{2}\left( \frac{1
}{4}\hat{\chi}^{2}-\hat{M}^{2}\right) +\frac{1}{4}\hat{\sigma}^{2}\hat{\chi}
^{2}\,\right\} \quad   \label{sugpot}
\end{equation}
and its respective derivatives.

The integration of the two coupled equations was performed by implementing a
variant of the Bulrish-Stoer$\cite{Rec}$ variable step method in a Fortran
program. As a general rule, the initial step for the dimensionless time
variable $\hat{t}$ was chosen to be 2 while the sought accuracy was put to $
10^{-15}$. This choice was found to ensure reasonable stability in all cases.

The results of our search are shown in Figs.1,2,3, matching the
corresponding sets of initial conditions, and 4 where supergravity
corrections were included in the set of initial conditions (1). Each point
on the $\hat{\sigma}$-$\hat{\chi}$ plane corresponds to given 
initial conditions and the way it is depicted corresponds to a definite
evolution type for the $\hat{\sigma}$-$\hat{\chi}$ system. We have used
three different kinds of symbols.

(a) Open circles: Both fields oscillate and fall towards the
supersymmetric minima without producing any appreciable amount of inflation.

(b) Filled triangles: Here, only $\hat{\chi}$ oscillates. The
field $\hat{\sigma}$ starts-off at large values ($\hat{\sigma}>\hat{\chi}$)
and remains almost constant for a very large period of time. The system
eventually relaxes at the bottom of the valley of minima of $\hat{V}(\hat{
\chi},\hat{\sigma})$ in eq.(\ref{reschub}). Its subsequent evolution along
this valley produces an adequate amount of inflation.

(c) Filled circles: Both fields start oscillating at the
beginning. Then $\hat{\sigma}$ settles down at large values and the system
follows an evolution of type (b) with adequate inflation.

\noindent The evolution pattern (b) includes the limiting area $\hat{\sigma}
\gg 1,\hat{\chi}$ already described and analyzed by means of semianalytic
arguments. In case (c), although $\hat{\sigma}$ starts mostly at 
moderate or small
values $\hat{\sigma}\thicksim $ $\hat{\chi}$, it appears to increase in
amplitude absorbing energy from the fast oscillating field $\hat{\chi}$,
creating eventually conditions favoring evolution of type (b). Here, we have
an example of large energy transfer between two strongly-coupled non-linear
oscillators. A closer look at Figs. 1-4 reveals that, in contrast to the
case of the smooth hybrid inflationary model, connected regions
corresponding to a definite evolution pattern do not appear to emerge except
for the filled triangles area in Fig.1. This area, however, although leading
to successful inflation cannot be considered as being ''natural'' since it
requires relatively large differences between the initial values of the
fields and is destroyed by supergravity as we shall soon see. Points
following evolution type (c) are significantly better since, for many
of them, the initial values of the fields are more or less of the 
same order of magnitude. These points,
however, are scattered and do not appear to constitute an extended and well
bounded continuous area as in the case of the smooth hybrid inflationary 
model. We have
checked in more detail the area around most of the filled circle points and
found out that almost all the points in their immediate neighborhood 
correspond to
the evolution  type (a) suggesting that evolution of type (c) occurs 
only for more or less isolated points.
We conclude then that, in the model analyzed here, only about 10\% of the
examined points lead to successful inflation and lie in the area which can 
be considered ''natural''.
Finally, the results with supergravity corrections
included (Fig.4) show that the filled triangle area disappears and we are
left with evolution of the types (a) and (b). This is to be expected, since
the very steep walls of the potential in the $\hat{\sigma}$ direction do not
allow the $\hat{\sigma}$ field to linger at large values for long. It is
remarkable to realize that here also only about 10\% of the examined points
lead to adequate inflation. The initial energy density $\rho $ can be
calculated from eqs. (\ref{resclq}),(\ref{reschub}) for global SUSY
or (\ref{resclq}),(\ref{sugpot}) in the supergravity case with $\kappa $
given by eq.(\ref{kappa}) where $\sigma _{\text{{\sc Q}}}/\sigma _{\text{
{\sc c}}}$ is taken close to $2$ \cite{LazSSh}.
It turns out that, for many initial conditions 
corresponding to type (c) evolution, we get adequately large initial energy
densities. In summary, only a scattered set of ''natural'' but more or
less isolated initial conditions, constituting about 10\% of the examined
cases, do lead to the SUSY inflationary model based on radiative
corrections. It appears that this scenario is less satisfactory than smooth
hybrid inflation which is obtainable for almost all initial conditions.
\acknowledgments{This work was supported in part by E.U. grant
 ERBFMRXCT 960090}

\newpage
\begin{center}
 \large{FIGURE CAPTIONS}
\end{center}
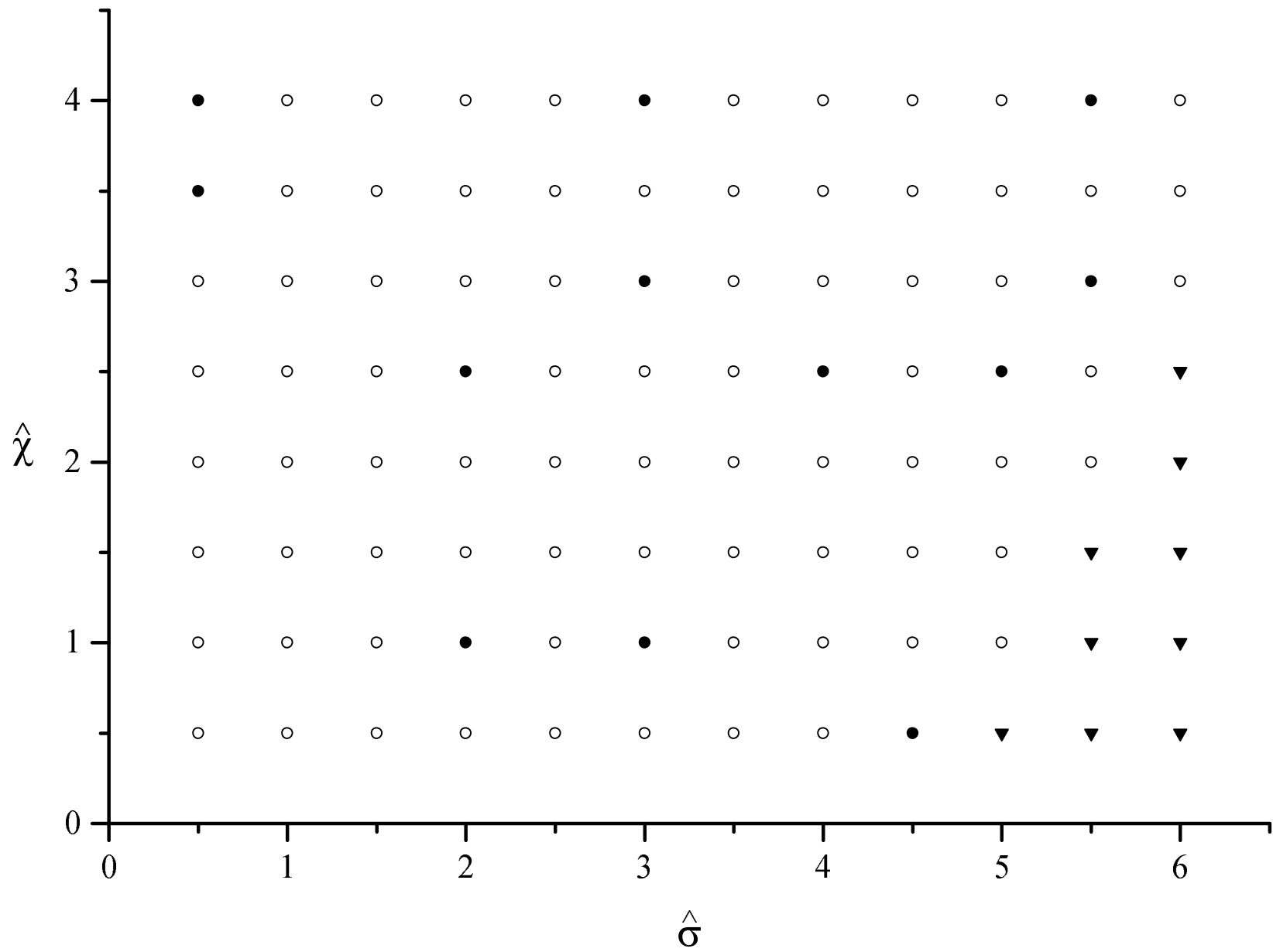
{\footnotesize \noindent {\bf Fig.~1}: Evolution patterns for the } $\hat{
\sigma}$-$\hat{\chi}$ {\footnotesize system with vanishing initial
velocities. Open circles represent points that do not lead to inflation.
Filled triangles give adequate inflation having only the } $\hat{\chi}$ 
{\footnotesize field oscillating. Filled circles give adequate inflation
with both fields oscillating initially. }

{\footnotesize \vspace{1cm} }

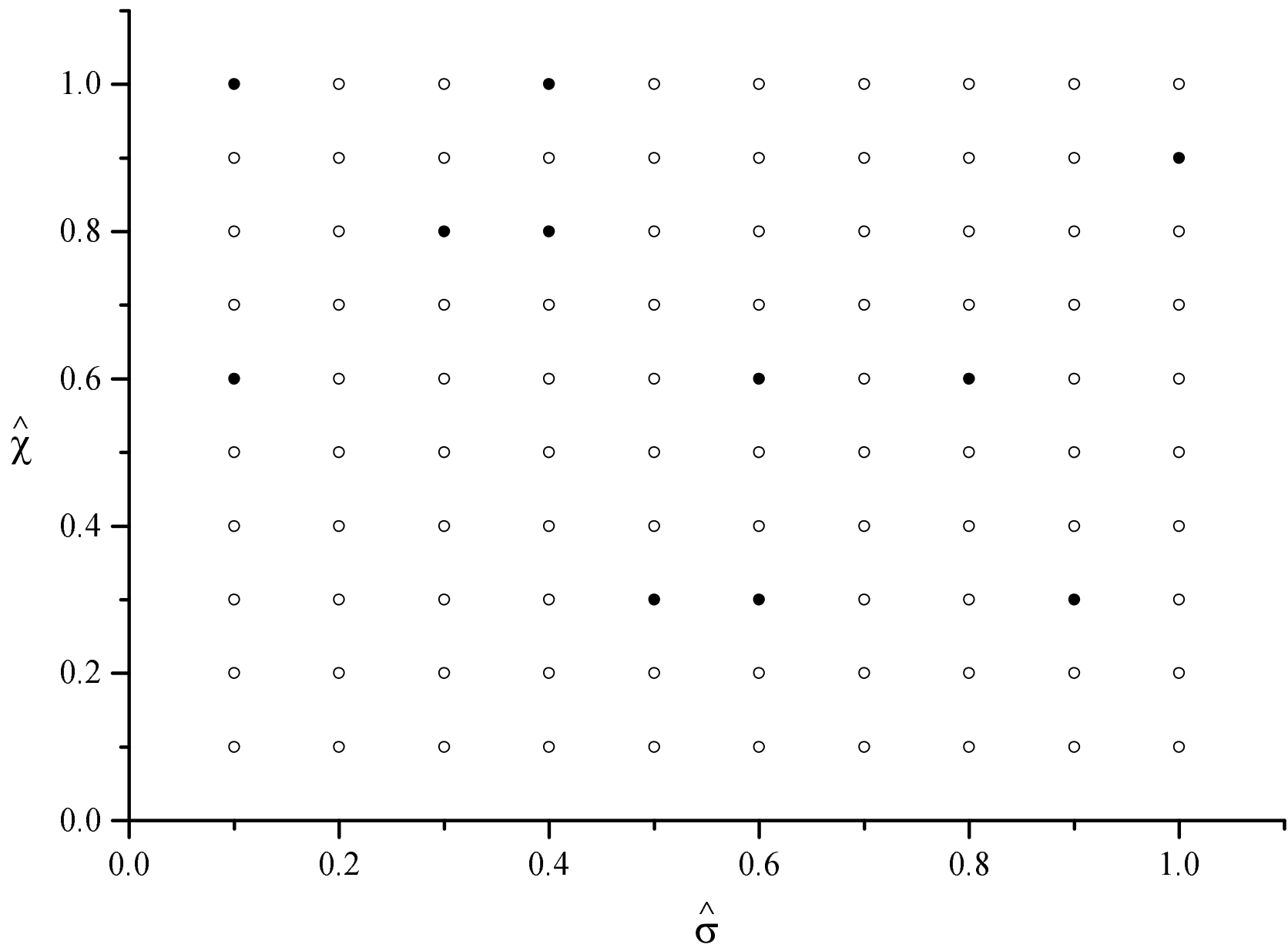
{\footnotesize \noindent {\bf Fig.~2}: Same as in Fig.1. The initial values
are now restricted to lie near the beginning of the axes. }

{\footnotesize \vspace{1cm} }

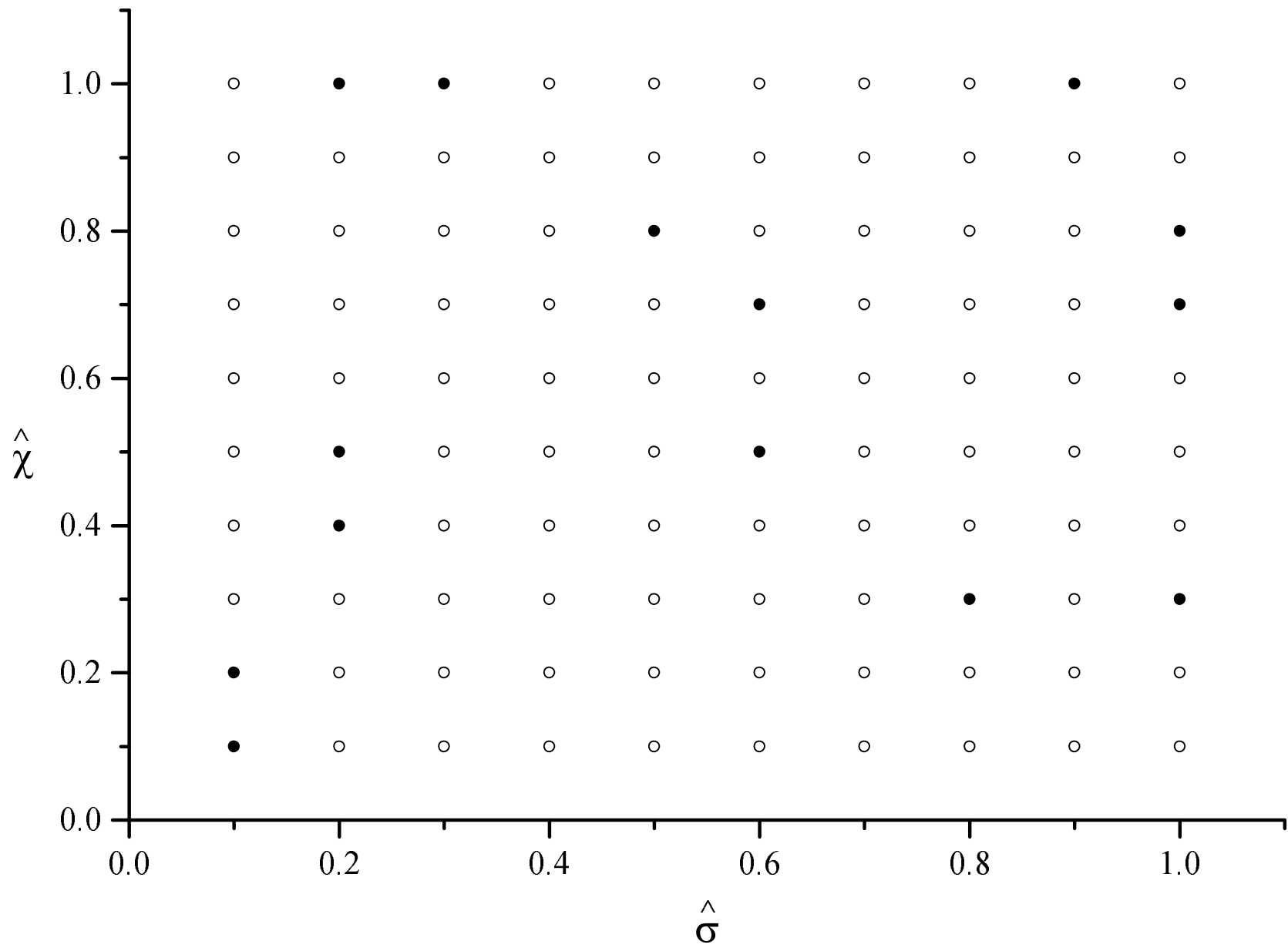
{\footnotesize \noindent {\bf Fig.~3}: Same as in Fig.2. with positive
initial velocities and equality of potential and kinetic energies.}

{\footnotesize \vspace{1cm} }

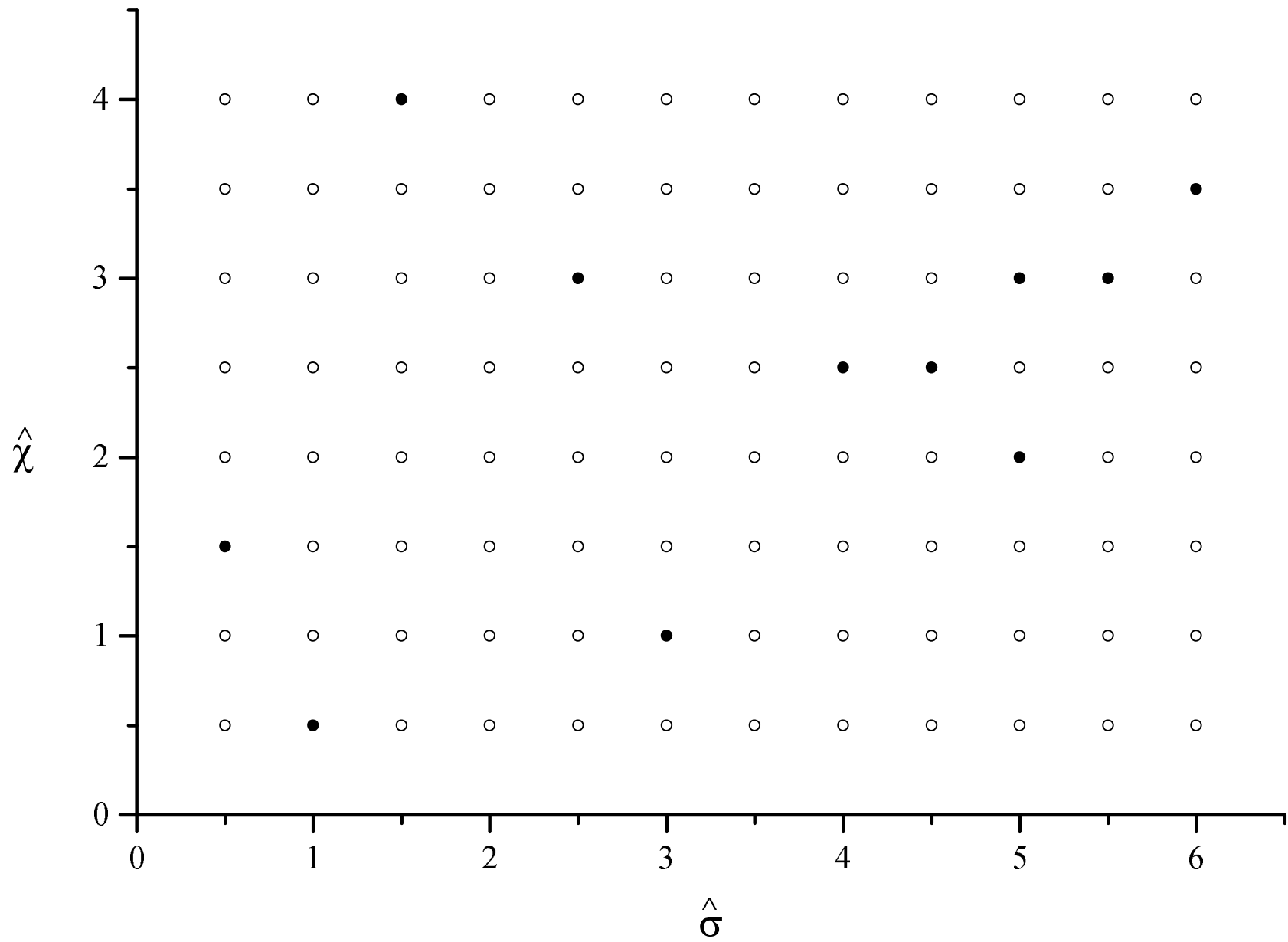
{\footnotesize \noindent {\bf Fig.~4}: Same as in Fig.1. Supergravity
corrections are now being taken into account.}

\end{document}